\documentclass[aps,preprint,onecolumn]{revtex4}
\usepackage{graphicx}

\begin{document}
\bibliographystyle{prsty}
\title{Surface plasmon propagation in an elliptical corral}
\author{A. Drezet, A. L. Stepanov, H. Ditlbacher, A. Hohenau, B. Steinberger, F. R. Aussenegg, A. Leitner, and J. R. Krenn}
\affiliation{Institut of Physics, Karl-Franzens University Graz, Universitätsplatz 5 A-8010 Graz, Austria }

\date{\today}

\begin{abstract}
We report the experimental realization of an elliptical Bragg reflector acting as an interferometer for
propagating surface plasmon sSPd waves. We investigate SP interferometry in this device using a
leakage radiation microscope and we compare our observations with a theoretical model for SP
propagation. Strong SP focalization as a function of laser polarization orientation is observed and
justified.
\end{abstract}

\maketitle

Interferometry is at the heart of wave optics. For a long
time interferometers were only realizable on the macroscale,
however, with the miniaturization of the optical technology,
the possibility of realizing microscale interferometers arose.
Recent developments towards two-dimensional (2D) optics
show that surface plasmons (SPs), which are quasi-2D electromagnetic
modes confined at the interface between a dielectic
and a metal \cite{1}, can interfere and generates fringes \cite{2,3}.
With SPs, unlike in conventional optics, it becomes imaginable
to realize highly integrated optical planar devices, allowing
one to overcome the classical diffraction limit. The
application of SPs to interferometry experiments with micrometer
sized devices has been recently exploited for demonstrating
a Mach–Zehnder configuration \cite{4}, which constituted
a first step for the implementation of SP wave optics. In this
letter, we extend this precedent work to demonstrating an
optical corral for SP waves, by using an elliptical Bragg reflector.
In such an elliptical corral, acting as an interferometer,
we generate stationary waves by launching a SP in one
focal point $F_1$, to be focused in the other one $F_2$. We describe
the observed interferences and SP focalization inside the elliptical
corral, and we compare our observations with a
simple numerical model.\\
In order to achieve a significant SP reflectivity and to
reduce SP scattering to light we used in Ref. 4 a mirror
composed of protrusions on a silver thin film, arranged in
parallel lines. To obtain a reflectivity better that 90\% we
worked in the Bragg regime \cite{5}, depending strongly on the SP
wavelength lSP, the separation distance between the protrusions
lines, and the incidence angle. However, the Bragg
conditions cannot be fulfilled for all directions that appear in
a diverging beam. In an elliptical corral, however, the length
of all SP paths connecting the two foci $F_{1,2}$ to any point on
the mirror equals $2a$, i.e., the long axis length. It is thus
straightforward to realize Bragg's condition by using confocal
ellipses. Constructive interference between the reflected
contributions of the different ellipses will be present only if
the variation $\delta(2a)$ of the long axis length between two concentric
ellipses is equal to $N\lambda_{SP}$, where $N$ is an integer. As
this condition is independent of angle the Bragg resonance is
easy to obtain even with divergent SP beams. This property
constitutes one of the essential motivations for elaborating an
elliptical interferometer.
The SP optical elements are obtained by the same electron
beam lithography method as described in Ref. 4. The
elements, as sketched in Fig. 1(a), are composed of $SiO_2$
protrusion (160 nm diameter, 80 nm height, center-to-center
distance 250 nm), deposited on a glass substrate and covered
by a layer of gold of 80 nm thickness. The protrusions are
arranged in five confocal ellipses, all of which share the
same two focal points, separated by 30 mm. For a SP wavelength
of $\lambda_{SP}\simeq750$ nm, accessible to a tunable Ti:sapphire
laser ($\lambda_0\simeq$ 750 nm)\cite{1}, we set $\delta a$ to 375 nm ($N=1$). The long
half axes of the confocal ellipses measure from $a_{min}
=30$ $\mu$m to $a_{max}=31.5$ $\mu$m and the number of protrusions
per ellipse varies from $n_{min}=832$ to $n_{max}=932$ [compare Fig.
1(a)]. For local SP launch, a single protrusion (160 nm diameter,
80 nm height) is fabricated at the position of $F_1$.
The measurement of the SP fields is based on a leakage
radiation (LR) microscope, sketched in Fig. 1(b). A linearly
polarized laser beam is focused with a microscope objective
$O_1$ (50$\times$, numerical aperture =0.7) onto the sample plane.
By focusing on the single protrusion in $F_1$ a SP wave is
launched\cite{6} in the corral. Due to the presence of the glass
substrate the SP is a leaky wave, giving rise to LR along the
cone defined by the angle $\theta_{LR}$ (which is always larger than
the critical angle of total internal reflection), fulfilling the
phase matching condition \cite{1,2,7}. As the LR intensity is proportional
to the SP intensity in any point of the plasmon carrying
interface, an immersion objective ($O_2$, 63$\times$, numerical aperture=
1.25) and a charged-coupled-device (CCD) camera can
be used to record leakage radiation images as maps of the SP
intensity distribution.\\
A SP wave generated by a subwavelength protrusion is
characterized by a $[\cos{(\theta)}]^2$ angular intensity distribution,
where $\theta$ is the azimuthal angle between the polarization axis
of the laser and the direction to the observation point. SP
reflection at the corral gives rise to a corresponding interference
pattern. By rotating the laser polarization with a half
wave plate we must then observe modifications of the interferences
in the elliptical corral. Figures 2(a) and 2(b) show
the interference patterns obtained for two polarization orientations
(indicated by the double sided black arrows), while
focusing the laser beam onto the single protrusion in $F_1$. We
see clearly the presence of fringes in the elliptical corral and
a strong focalization of the SP beam at $F_2$. We note that the
grid of white spots in Figs. 2(a) and 2(b) is an artifact from
the CCD camera, and that part of the images has been saturated
to increase the visibility of the interference pattern.
To theoretically model the SP wave we assume at every
point on the sample plane the sum of a direct SP $\Psi_I$ launched
from $F_1$ and of a reflected  SP $\Psi_R$ coming from the individual
protrusions of the mirror. The field is calculated in the first
Born approximation \cite{8} and takes into account the
$\cos{(\theta)}e^{ik_{SP}r} /\sqrt{r}$ dependance \cite{2,6,9} for the source term ($k_{SP}$ is
the complex SP in-plane wave vector and $r$ the distance to
the observation point). The response of each individual protrusion
is supposed to be proportional to the local SP intensity
and is further assumed isotropic with the same radial
dependence as $\Psi_I$. As in Ref. 4 a maximum Bragg mirror
reflectivity of 90\% is found experimentally, we introduce this
value as the only free parameter into the model. The thereby
calculated SPs profile in Figs. 2(c) and 2(d) fit the experiment
quite well.\\
In a next step, we decrease the exiting intensity to prevent
saturation in the images and thereby allow for quantitative
measurement. In Figs. 3 and 4, we compare experiment
and theory for different polarization angles $\theta$ of the exiting
laser beam and, again, we find a good qualitative agreement.
From the experimental data we extract the SP intensity in $F_2$
and plot it in Fig. 5 as a function of the polarization angle $\theta$.
Due to the angular dependence of the excitation, $\Psi_I$ must
follow a $cos{(\theta)}$ law. The same is true for $\Psi_R$ due to interference
of the SPs reflected by the protrusions. It is worth
mentioning that these interferences come essentially from the
angular dependence of the reflectivity of the mirror \cite{10} and not
from the complex phase factor $e^{ik_{SP}L}$ which is constant for every SP light path of length $L=2a$ connecting $F_1$ and $F_2$ to
any points on the mirror. The total intensity then obeys a
$[\cos{(\theta)}]^2$ law which is perfectly reproduced \cite{11} by the experimental
data extracted from Fig. 3 and plotted in Fig. 5.\\
In summary, we have elaborated an elliptical Bragg reflector
acting as a corral for SP waves. SP propagation is
observed using a LR microscope which allows quantitative
observations and comparison with a theoretical model. By
launching a SP wave at $F_1$ we observe numerous fringes and
a strong focalization in $F_2$. We can finally realize an angular
interferometer by rotating the polarization of the exciting
laser beam. In all cases the observations are in qualitative as
well as quantitative agreement with our model.\\
For financial support the Austrian Science Foundation
and the European Union, under Project Nos. FP6 NMP4-CT-
2003-505699 and FP6 2002-IST-1-507879, are acknowledged

\begin{figure}[h]
\includegraphics[width=10cm]{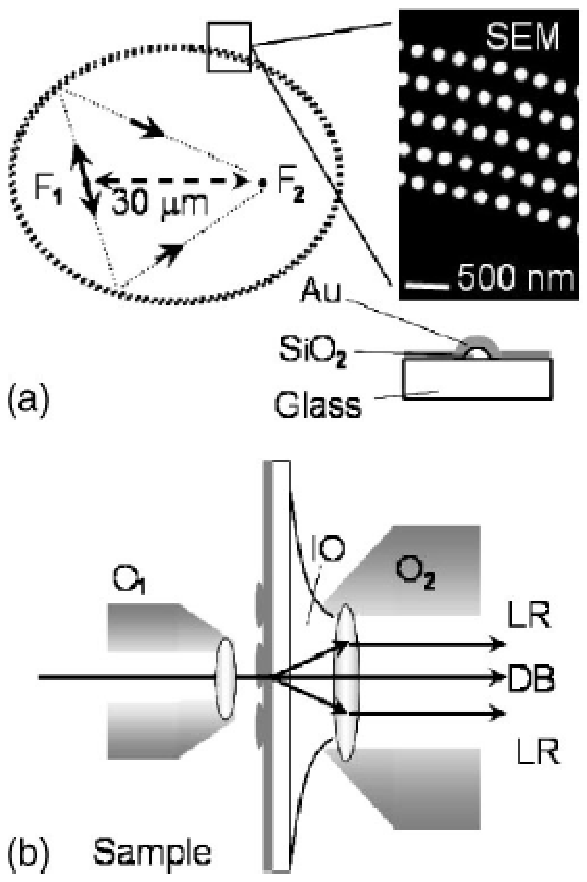}
\caption{(a) Outline of the elliptical interferometer together with a scanning
electron microscopy image of the mirror structure and a sketch of an individual
protrusion. $F_{1,2}$ foci of the ellipse. (b) Sketch of the leakage radiation
microscope. $O_{1,2}$ microscope objectives, LR leakage radiation, DB direct
beam, IO immersion oil.}
\end{figure}
\begin{figure}[h]
\includegraphics[width=10cm]{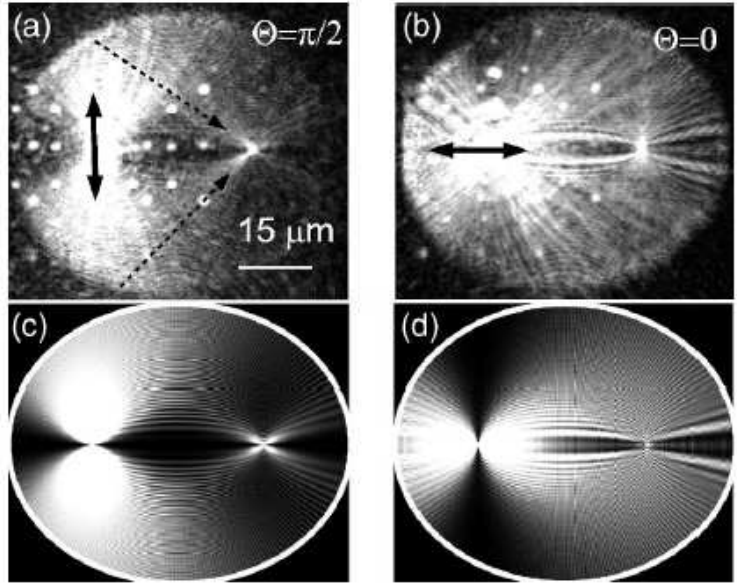}
\caption{LR imaging of SP modes propagating in the elliptical corral for (a)
vertical ($\theta=\pi/2$) and (b) horizontal polarizations $(\theta=0)$ and $\lambda_0\simeq 750$ nm.
The regular grid of white spots is an artifact from the CCD camera. (c) and
(d) The corresponding simulations based on the model.}
\end{figure}
\begin{figure}[h]
\includegraphics[width=10cm]{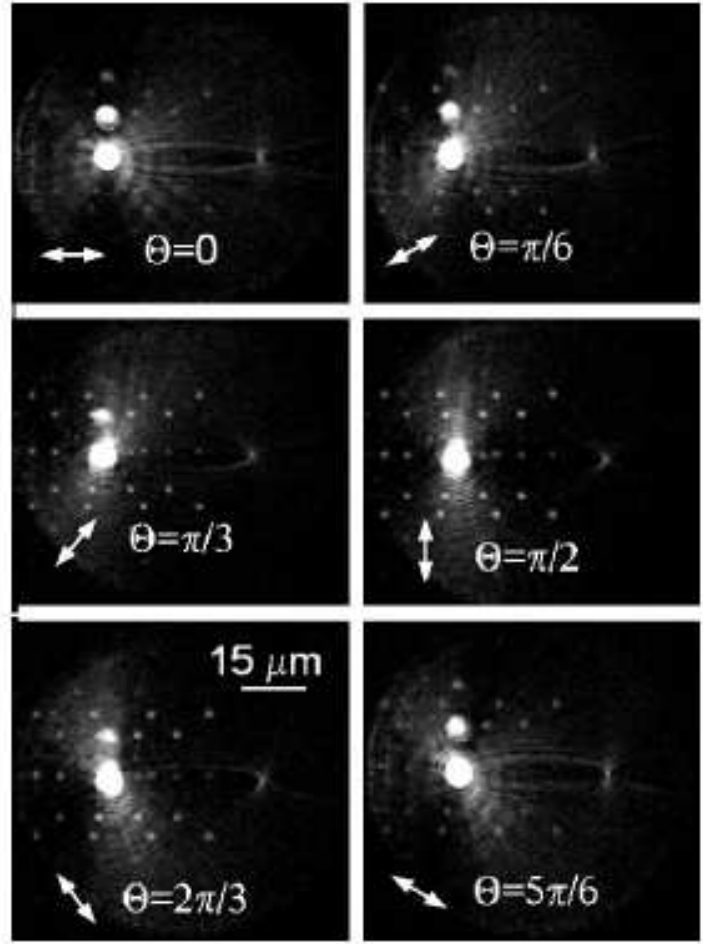}
\caption{LR images of SPs wave propagating from $F_1$ to $F_2$ for different
polarization angles ($\lambda_0\simeq 750$ nm). The polarization directions are indicated
by the double sided white arrows. The regular grid of white spots is an
artifact from the CCD camera.}
\end{figure}
\begin{figure}[h]
\includegraphics[width=10cm]{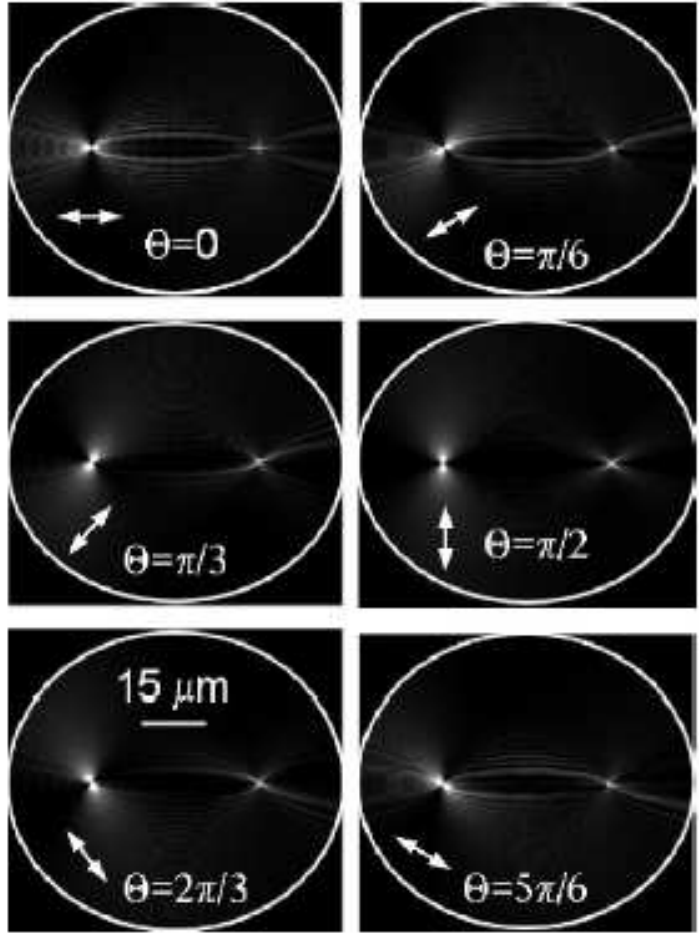}
\caption{Simulation of the LR images of Fig. 3. The polarization directions
are indicated by the double sided white arrows.}
\end{figure}
\begin{figure}[h]
\includegraphics[width=10cm]{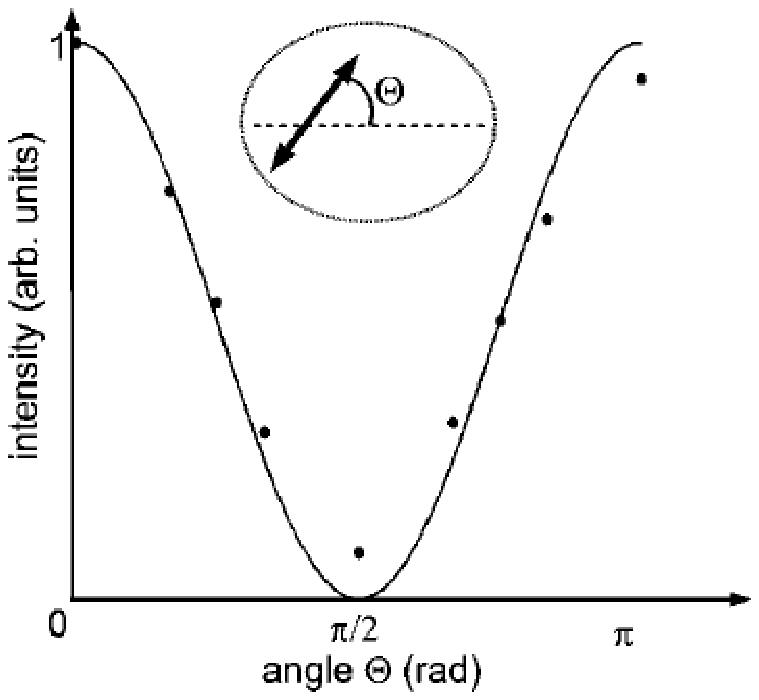}
\caption{Experimental data (dots) and theoretical model (solid line) for the
dependence of the SP intensity at $F_2$ on the polarization angle $\theta$.}
\end{figure}


\begin{references}
\bibitem{11} The $[\cos{(\theta)}]^2$ law is in fact valid for every point located on the long
ellipse axis. While the intensity at $F_2$ equals zero theoretically (as can be
observed by zooming on this point in the simulation) the LR images in
Fig. 3 do not have sufficient spatial resolution to observe this effect. Consequently,
the experimental data in Fig. 5 represent the average of the data
points around $F_2$.
\end{references}
\end{document}